\documentclass[english,twocolumn,prl,amssymb,amsmath,superscriptaddress,showpacs]{revtex4}
\usepackage[latin9]{inputenc}
\usepackage{graphicx,color}
\usepackage[bookmarks]{hyperref}
\usepackage{babel}

\newcommand{\be}{\begin{equation}}
\newcommand{\ee}{\end{equation}}
\newcommand{\ua}{\uparrow}
\newcommand{\da}{\downarrow}

\begin{document}

\title{Dynamics of symmetry breaking during quantum real-time evolution in a minimal model system}

\author{Markus Heyl}
\affiliation{Institute for Quantum Optics and Quantum Information of the Austrian Academy of Sciences, 6020 Innsbruck, Austria}
\affiliation{Institute for Theoretical Physics, University of Innsbruck, 6020 Innsbruck, Austria}
\author{Matthias Vojta}
\affiliation{Institut f\"ur Theoretische Physik, Technische Universit\"at Dresden, 01062 Dresden, Germany}

\begin{abstract}

One necessary criterion for the thermalization of a nonequilibrium quantum many-particle system is ergodicity. It is, however, not sufficient in case where the asymptotic long-time state lies in a symmetry-broken phase but the initial state of nonequilibrium time evolution is fully symmetric with respect to this symmetry. In equilibrium one particular symmetry-broken state is chosen due to the presence of an infinitesimal symmetry-breaking perturbation. We study the analogous scenario from a dynamical point of view: Can an infinitesimal symmetry-breaking perturbation be sufficient for the system to establish a nonvanishing order during quantum real-time evolution? We study this question analytically for a minimal model system that can be associated with symmetry breaking, the ferromagnetic Kondo model. We show that after a quantum quench from a completely symmetric state the system is able to break its symmetry dynamically and discuss how these features can be observed experimentally.

\end{abstract}
\date{\today}
\maketitle

At the heart of statistical mechanics and thermodynamics lies the assumption that realistic macroscopic
physical systems exhibit one particular state - thermal equilibrium - that is always approached irrespective
of the initial condition. From a fundamental point of view the important question, however, which microscopic conditions are necessary or sufficient for the thermalization of a \emph{closed} quantum system is still largely unanswered~\cite{Polkovnikov2011kx}. This is of particular importance especially because there exists a specific class of isolated quantum systems termed integrable for which equilibration is hindered by the presence of special conservation laws as demonstrated experimentally in the quantum version of Newton's cradle~\cite{Kinoshita2006qg}. Generally, it is believed that the class of so-called nonintegrable systems can thermalize because they are complex enough in order to be ergodic~\cite{Polkovnikov2011kx}. 

Ergodicity, however, is not always sufficient for thermalization even though the system under study may be nonintegrable. This is the case whenever the asymptotic long-time state lies in a symmetry broken phase but the initial state is fully symmetric. As the Hamiltonian conserves this symmetry by construction, the system can never break this symmetry by itself, but rather requires some symmetry-breaking perturbation from the exterior. In this work we address the fundamental question whether an infinitesimal symmetry-breaking perturbation can establish order dynamically during nonequilibrium quantum real-time evolution. Clearly, in case where the order parameter itself is a constant of motion this is impossible. Here we demonstrate that for the opposite case a symmetry breaking can occur dynamically.

From a theoretical point of view, symmetry breaking in equilibrium is a consequence of the noncommutativity of two limits: $\lim_{h\to0}\lim_{L\to\infty} \not=\lim_{L\to\infty}\lim_{h\to0} $ where $h$ refers to the strength of the symmetry-breaking perturbation (a magnetic field, for example) and $L$ to the system size. From a dynamical perspective, we propose the following criterion for the breaking of a symmetry during real-time evolution: the noncommutativity of two different limits $\lim_{h\to0}\lim_{t\to\infty} \not=\lim_{t\to\infty}\lim_{h\to0} $ with $t$ time (and system size $L\to \infty$ by default). Phrased differently, is it possible that a magnetic field is capable to establish magnetic order in a quantum magnet after a sufficiently long time even though its strength may be arbitrarily small? Importantly, the influence of the symmetry-breaking field cannot be treated perturbatively such that it is necessary to study the full time evolution of the quantum many-body system into the symmetry-broken phase.

In classical systems, the buildup of ordered structures out of metastable disordered states, e.g., crystallization of undercooled liquids, has been studied extensively~\cite{Binder1987}. These metastable states may either develop instabilities, such as in the context of spinodal decomposition, or decay into the respective stable thermodynamic equilibria via the formation of droplets in case of nucleation. Classical nucleation is driven by thermal fluctuations induced by a surrounding bath, but the transition from metastable to stable states may also be induced by quantum fluctuations overcoming the potential barrier between the two states via quantum tunneling~\cite{Coleman1977}. Here, we will be interested in the buildup of order during unitary real-time evolution in a minimal quantum magnet far beyond equilibrium where no notion of metastable states in free-energy landscapes exists. Recently, the buildup of antiferromagnetic order in the Hubbard model has been investigated for cases where the symmetry-breaking perturbation acts over finite time intervals~\cite{Tsuji2013}. Moreover, for the Lieb-Mattis model it has been shown that the symmetry-breaking perturbation is not capable of inducing nonzero order except at a periodic sequence of singular points in time ~\cite{Ortix2011}.

\emph{Ferromagnetic Kondo model:-} We will demonstrate exemplary the anticipated ideas for a minimal model system, the ferromagnetic Kondo model which by a quantum-classical mapping is equivalent to the one-dimensional $1/r^2$-Ising chain~\cite{Yuval1970}. In equilibrium, the $1/r^2$-Ising chain hosts a symmetry-broken phase at low temperatures with a nonzero magnetization~\cite{Froehlich1982} which is triggered by an infinitesimal magnetic field. We study the dynamics of symmetry breaking for the quantum system where the symmetry breaking is associated with a boundary quantum phase transition~\cite{Vojta2006} with the expectation that the main features observed are of generic relevance beyond the chosen model system. The ferromagnetic Kondo model
\begin{align}
      H_\mathrm{fKM} = \sum_{k,\sigma = \uparrow,\downarrow} &  \varepsilon_k c_{k\sigma}^\dag c_{k\sigma} + \frac{J}{2} \sum_{kk'} \left[ c_{k\ua}^\dag c_{k'\ua} - c_{k\da}^\dag c_{k'\da} \right] S^z \nonumber \\+ & \frac{J}{2} \sum_{kk'} \left[ c_{k\ua}^\dag c_{k'\da} S^- + c_{k\da}^\dag c_{k'\ua} S^+ \right]
\label{eq:ferromagneticKondoModel}
\end{align}
describes a local spin-$1/2$ degree of freedom coupled via a ferromagnetic ($J<0$) exchange to a fermionic bath. For the following it is suitable to introduce the dimensionless coupling constant $g = \rho J$ with $\rho$ the noninteracting conduction band density of states which can be chosen constant within a band $[-D,D]$ for the universal properties of the model~\cite{Hewson1997}.

In equilibrium, the spin-$1/2$ becomes asymptotically free at low energies. Under a perturbative renormalization group (RG) transformation the dimensionless coupling constant obeys the following scaling equation at low energies~\cite{Hewson1997}
\be
      g(\Lambda) = \frac{g}{1+g \log(\Lambda/D)}.
\label{eq:JScaling}
\ee
The reduction of the UV-cutoff $\Lambda$ well below the electronic bandwidth $D$ leads to a logarithmic decay of  $g$. The fixed point is thus a free theory of an isolated spin decoupled from the fermionic bath. The free spin shares a rotational symmetry that is broken by any infinitesimal local magnetic field at zero temperature forcing an alignment along the magnetic field direction. This yields a local magnetization equal to~\cite{Abrikosov1970}
\be
	\langle S_z \rangle = \frac{1}{2} \left[ 1+\frac{g}{2}+\mathcal{O}\left( g^2 \right) \right]
\label{eq:magnetizationEquilibrium}
\ee
for a small bare coupling $g$. Notice that the low-energy properties of the \emph{antiferromagnetic} Kondo model with $J>0$ differ on a fundamental level. The system flows to strong instead of weak coupling, leading to a Kondo singlet ground state~\cite{Hewson1997} which does not exhibit symmetry breaking.

\begin{figure}
\centering
\includegraphics[width = 0.7\columnwidth]{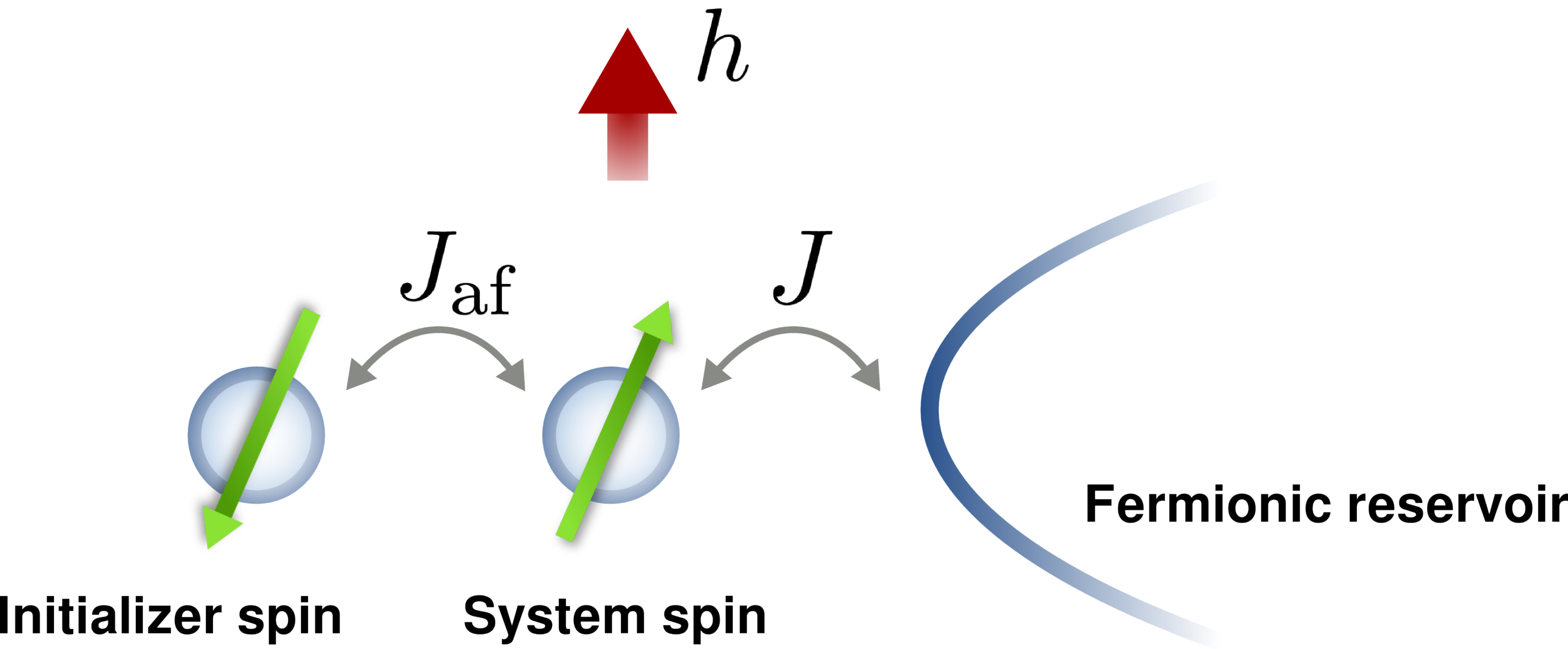}
\caption{(color online) Dynamics of symmetry breaking in the ferromagnetic Kondo model. The proposed setup is composed out of three elementary components. The system spin and the fermionic reservoir, coupled via a ferromagnetic exchange $J$, are supposed to realize the ferromagnetic Kondo model. During the dynamics of symmetry breaking in presence of an infinitesimal local magnetic field $h$ the system spin develops a local magnetization. The antiferromagnetic coupling $J_\mathrm{af}$ to the initializer spin generates an initially rotationally symmetric spin singlet.}
\label{fig:1}
\end{figure}

We develop a dynamical theory for symmetry breaking in the ferromagnetic Kondo model. Our setup is illustrated in Fig.~\ref{fig:1}. It consists of two antiferromagnetically coupled magnetic moments of spin $1/2$, one of them we call the initializer spin, the other the system spin. Additionally, the system spin is also coupled to an electronic environment through a ferromagnetic exchange interaction implementing the Hamiltonian in Eq.~(\ref{eq:ferromagneticKondoModel}). Such a ferromagnetic exchange can be realized in specific designs of triple quantum dot systems where it has been shown that it is possible to obtain effective ferromagnetic Kondo models, either anisotropic~\cite{Kuzmenko2006,Roosen2008,Mitchell2009} or isotropic~\cite{Baruselli2013}. Notice that the initializer spin is not coupled to the electronic reservoir.

We initialize a rotationally symmetric state by decoupling the system spin from the electronic reservoir in presence of the antiferromagnetic coupling leading to a spin singlet of the two local magnetic moments. This can be achieved by choosing $J_\mathrm{af}$ as the largest energy scale in the problem. After this initialization procedure, we switch off the antiferromagnetic coupling, inducing nonequilibrium real-time dynamics for the system spin according to the Hamiltonian in Eq.~(\ref{eq:ferromagneticKondoModel}) while the initializer spin is decoupled from the dynamics.

As the symmetry-breaking perturbation we add a (infinitesimally) small local magnetic field $h$ that couples to the system spin yielding as the full Hamiltonian
\be
	H = H_\mathrm{fKM} + H_h, \,\,\, H_h = -h S_z.
\label{eq:fullHamiltonian}
\ee
Both the Bohr magneton and the magnetic moment's $g$-factor have been absorbed into the definition of the magnetic field $h$. In the weak-coupling limit $|g|\ll D$, our setup possesses the following hierarchy of energy scales: $|h| \lesssim |g|\ll D$. In the following we will always choose $h>0$ for simplicity. 

We study the time evolution of the local magnetization $\langle S_z(t) \rangle$ driven by the Hamiltonian $H$. First, we will focus on the zero-temperature limit that hosts the symmetry-broken phase of the model. Later we will also discuss the nonzero-temperature case which is of particular importance for any experimental realization. According to the anticipated protocol, the initial state $\rho_\mathrm{o} = \rho_S \otimes \rho_B$ factorizes into the singlet $\rho_S = |S\rangle \langle S|$ of initializer and system spin and the Fermi sea $\rho_B$ of the electronic reservoir. 

\begin{figure}[t]
\centering
\includegraphics[width = \columnwidth]{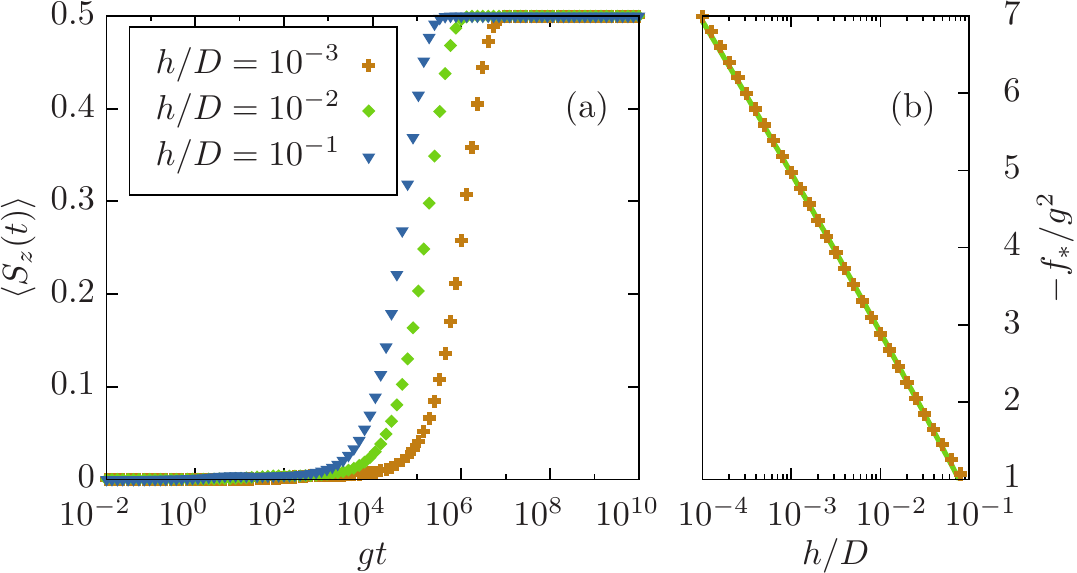}
\caption{(color online) (a) Dynamics of the local magnetization in the ferromagnetic Kondo model for different magnetic field strengths obtained from the numerically exact solution of the 1-loop flow equations for $g=10^{-2}$. For times $t<t^\ast$ the local magnetization only acquires perturbative corrections in presence of a small magnetic field. Beyond the time scale $t^\ast$, the magnetic field induces a local magnetic moment that saturates to a value independent of the magnetic field in the asymptotic long-time limit up to corrections that vanish in the zero field limit. (b) Asymptotic long-time value of the local magnetization $\langle S_z(t\to\infty)\rangle = (1+f_\ast)/2$. Comparison of the full numerically exact result for $f_\ast$ based on the 1-loop flow equations (points) to the analytical estimate $f_\ast/g^2 = 3/2+\log(h/D)/(1+g\log(h/D))$ (line) demonstrating the accuracy of both the numerical as well as analytical result.}
\label{fig:2}
\end{figure}

\emph{Results.-} We study the real-time dynamics analytically using the flow-equation technique~\cite{Kehrein2006} that has been proven to provide a very accurate description for the out of equilibrium dynamics in the ferromagnetic Kondo model. As has been shown in comparison to numerically exact time-dependent numerical renormalization group (TD-NRG) data, the flow-equation technique becomes asymptotically exact in the weak-coupling limit with well-controlled corrections for larger couplings~\cite{Hackl2009fk}.

We find that an infinitesimally small magnetic field establishes a time scale for dynamical symmetry breaking
\be
	t^\ast = \frac{\log^2(D/h)}{h}
\label{eq:tast}
\ee
that differs from a perturbative guess $h^{-1}$ by a large logarithmic factor indicating its nonperturbative influence onto the system's properties. For times $t\ll t^\ast$, the dynamics resembles the symmetric limit with $h=0$ up to perturbative corrections that vanish in the zero-field limit. For times $t \gg t^\ast$ the local moment develops a magnetization
\be
	\langle S_z(t) \rangle \stackrel{t\gg t^\ast}{\longrightarrow} \frac{1}{2} \left[ 1+g  \right]
\label{eq:magnetizationDSB}
\ee
whose magnitude is independent of the field strength $h$. The infinitesimal magnetic field $h$ breaks the rotational symmetry and forces the system to develop a nonvanishing magnetization.  Consequently, the limits $\lim_{t\to\infty}$ and $\lim_{h\to0}$ do not commute demonstrating dynamical symmetry breaking in a quantum many-body system. Notice that the asymptotic magnetization is not thermal, compare Eq.~(\ref{eq:magnetizationEquilibrium}) and Refs.~\cite{Hackl2009fk,Hackl2009,Pletyukhov2010}.
It is important to note, however, that thermalization is not relevant for dynamical symmetry breaking which only relies on the noncommutativity of the limits $\lim_{t\to\infty}$ and $\lim_{h\to0}$.

\emph{Zero temperature:-} The flow equation approach is an RG scheme under whose RG flow the Hamiltonian becomes more and more energy-diagonal successively~\cite{Kehrein2006}. This is done by constructing explicitly a unitary transformation $U(B)=\mathcal{T}_B \exp[\int_0^B dB \eta(B)]$ as a $B$-ordered exponential of its generator $\eta(B)$ and an associated family of Hamiltonians $H(B)=U(B) H U^\dag(B)$. For $B=0$ one recovers the initial Hamiltonian while for $B\to\infty$ the Hamiltonian becomes diagonal in energy and exactly solvable.

For the calculation of the magnetization we introduce a novel scheme for evaluating observables within the flow equation framework that avoids the separate solution of an additional set of scaling equations for the respective observables at the same level of accuracy~\cite{supp}. Instead, we utilize explicitly the exponential structure of the diagonalizing unitary transformation $U(B)$ by performing an operator cumulant expansion~\cite{Kubo1962xh} as has been done in the case of the Loschmidt echo~\cite{Heyl2012}. This yields for the magnetization~\cite{supp}
\be
	\langle S_z(t) \rangle = \frac{1}{2} \Big[ e^{f_\ua(t)} - e^{f_\da(t)} \Big]
\ee
with $f_{\sigma}(t) = \int d\varepsilon d\varepsilon' \mathcal{J}^2_{\varepsilon\varepsilon'} N_{\varepsilon\varepsilon'}(\sigma) \{1-\cos[(\varepsilon'+h^\ast )t] \}  $, $\mathcal{J}_{\varepsilon\varepsilon'} = \int_0^\infty dB \,\, g_\varepsilon^\perp(B) [\varepsilon'+h(B)] e^{-B(\varepsilon'+h(B))^2}$, and $N_{\varepsilon \varepsilon'}(\sigma) = n_{\varepsilon-\sigma\varepsilon'}(1-n_{\varepsilon+\sigma\varepsilon'})$ with $n_\varepsilon$ the Fermi-Dirac distribution and  $\sigma=\pm 1/2$ for $\sigma=\ua,\da$. Under the RG transformation by increasing the flow parameter $B$ the magnetic field $h(B)$ renormalizes yielding a $B$ dependence and asymptotically for $B\to\infty$ reaches a final value $h^\ast = h(B\to\infty)$. The dimensionless couplings $g$ develop an energy dependence under the flow and the presence of a magnetic field additionally introduces an anisotropy~\cite{supp}. For the magnetization only the renormalized spin flip coupling $g^\perp_\varepsilon(B)$ enters. In Fig.~\ref{fig:2} the results for the dynamics of the magnetization at zero temperature are shown and compared to the analytical estimates that will be presented below.

On intermediate time scales $D^{-1} \ll t \ll h^{-1}$, the two spin contributions $f_{\ua}(t)=f_{\da}(t)= f_\ast$ are identical such that $\langle S^z(t) \rangle = 0$ up to perturbative corrections. Thus, the symmetry-breaking perturbation is not capable to induce a local spin polarization in this regime. On these intermediate time scales we obtain the analytical estimate $f_\ast = g^2[1+2\log(h/D)/(1+g\log(h/D))]/2$~\cite{supp} for $h>De^{1/g}$ and $f_\ast = g$ for $h<D^{1/g}$ which are continuously connected. In Fig.~\ref{fig:2} we compare these analytical predictions to the numerically exact solution of the one-loop flow equations showing perfect agreement. Notice that for a fully polarized local initial state without rotational symmetry we would have $\langle S_z(t) \rangle = e^{f_\ast}/2= (1+g + \mathcal{O}(g^2))/2$ which is precisely the result obtained in previous works~\cite{Hackl2009fk,Hackl2009,Pletyukhov2010}, confirming the accuracy of the current calculation.

For times $t\gg t^\ast$, compare Eq.~(\ref{eq:tast}), the spin-$\ua$ component $f_\ua(t) = f_\ast = \mathrm{const.}$ is frozen while the spin-$\da$ component $f_\da(t) = -t/t_h$ shows a linear divergence. Thus, asymptotically for large times the local spin develops a magnetization exponentially fast~\cite{supp}
\be
	\langle S^z(t) \rangle \stackrel{t\gg t^\ast}{\longrightarrow} \frac{1}{2} \Big[ 1+f_\ast \Big] - \frac{1}{2} \Big[ 1+f_\ast \Big] e^{- t/t_h},
\ee
with a relaxation time
\be
	t_h = \sqrt{\frac{8}{\pi}} \left[\frac{1+g\log(h/D)}{g} \right]^2 \frac{1}{h}.
\label{eq:Gamma}
\ee
In the zero-field limit, one obtains to leading order $t_h \to \sqrt{8/\pi}\,\, t_\ast$ yielding the desired result in Eq.~(\ref{eq:tast}). Notice the relation to the linear-response spin relaxation rate $\Gamma \propto t_\ast^{-1}$~\cite{Goetze1971}.

\emph{Nonzero Temperatures:-} In equilibrium, the local magnetization in the ferromagnetic Kondo model is nonvanishing only at zero temperature. This poses a severe challenge onto the possibility to observe the anticipated dynamical symmetry breaking in experiments that necessarily operate at nonzero temperatures. We will demonstrate below that although any nonzero temperature will eventually lead to a completely symmetric state with vanishing magnetization, this will happen only beyond a time scale $t_T$. Thus, on intermediate times $ t<t_T$ it is possible to observe the dynamical symmetry breaking provided temperature is sufficiently small such that $t_T \gtrsim t_h$, see Eq.~(\ref{eq:Gamma}). Notice that temperature not only leads to a broadening of the Fermi-Dirac distribution, but also influences the preparation of the initial local singlet of initializer and system spin. In order to observe the desired dynamics it is therefore important to choose the antiferromagnetic coupling $J_\mathrm{af}\gg T$, such that temperature will only lead to exponentially small corrections in the initial state preparation.
\begin{figure}
\centering
\includegraphics[width=\columnwidth]{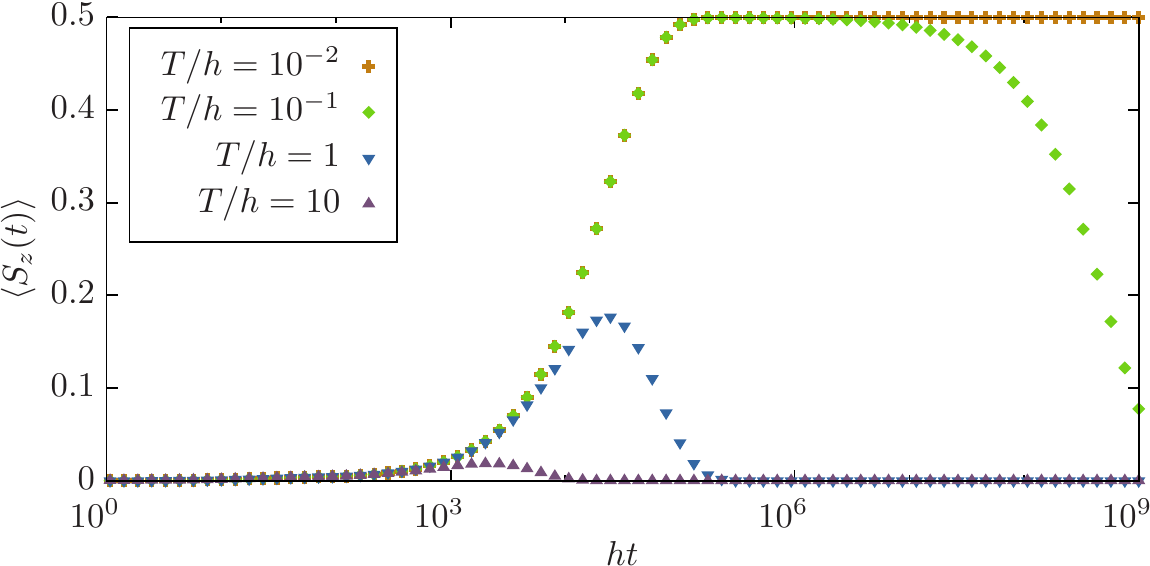}
\caption{(color online) Influence of temperature onto dynamical symmetry breaking in the ferromagnetic Kondo model. Dynamics of the local magnetization for different temperatures with $g=10^{-2}$ and $h/D=10^{-4}$. For small times $t<t_T$, see Eq.~(\ref{eq:tT}) the real-time evolution of the magnetization is equivalent to the zero-temperature limit demonstrating the observability of dynamical symmetry breaking in the ferromagnetic Kondo model at nonzero temperatures. Increasing temperature such that $t_T\lesssim t_\ast$ destroys the signatures of symmetry breaking eventually leading to a complete suppression in the limit $T\gg h$.}
\label{fig:3}
\end{figure}

While to leading order any nonzero temperature will not influence the properties of $f_\da(t)$, its influence onto the spin-$\ua$ component $f_\ua(t)$ is substantial for times $t\gg t_T$ with
\be
	t_T = \left[\frac{1+g\log(h/D)}{g} \right]^2  \frac{e^{h/T} -1}{\pi h},
\label{eq:tT}
\ee
where $f_\ua(t)=- t/t_T$ such that temperature induces an exponential decay of the magnetization
\be
	\langle S_z(t)\rangle \stackrel{t\gg t_T}{\longrightarrow} \frac{1}{2} \Big[ 1+f_\ast \Big] e^{-t/t_T},
\ee
reestablishing a completely symmetric state. For $h\gg T$ we have that $t_T\sim e^{h/T}$  yielding a symmetry-broken magnetization plateau which becomes stabilized to exponentially long times. This changes as soon as $h\lesssim T$ where $t_T\sim T^{-1}$ and $t_T\lesssim t_h$ preventing the buildup of a substantial magnetization plateau. In Fig.~(\ref{fig:3}), the real-time evolution of the magnetization and its dependence on temperature is shown, confirming the analytical arguments.

\emph{Conclusion:-} From a thermodynamical perspective any realistic system with a nonequilibrium initial condition evolves towards one particular state - thermal equilibrium -  regardless of its microscopic details. In case where the final state lies in a symmetry-broken phase the system has to break the associated symmetry dynamically during nonequilibrium real-time evolution provided the initial condition is compatible with this symmetry. In this work we have demonstrated this concept for a minimal quantum many-body system, the ferromagnetic Kondo model. Based on the analytical solution of the long-time real-time evolution we showed that  the system develops a nonzero local magnetization even in the limit where the symmetry-breaking magnetic field $h$ is infinitesimally small implying the noncommutativity of the two limits $h\to0$ and time $t\to\infty$.

\begin{acknowledgments}
This work has been supported by the DFG FOR 960 and GRK 1621, by GIF through 
grant G 1035-36.14/2009, and by the Deutsche Akademie der Naturforscher Leopoldina under grant number LPDS 2013-07.
\end{acknowledgments}

%%%%%%%%%%%%%%%%%%%%%%%%%%%%%%%%%%%%%%%%%%%%%%%%%%%%%%%%%%%%%%%%%%%%%%%%%%

\bibliographystyle{apsrev}
\bibliography{literatureDSB}

%%%%%%%%%%%%%%%%%%%%%%%%%%%%%%%%%%%%%%%%%%%%%%%%%%%%%%%%%%%%%%%%%%%%%%%%%%
%%%%%%%%%%%%%%%%%%%%%%%%%%%%%%%%%%%%%%%%%%%%%%%%%%%%%%%%%%%%%%%%%%%%%%%%%%
%%%%%%%%%%%%%%%%%%%%%%%%%%%%%%%%%%%%%%%%%%%%%%%%%%%%%%%%%%%%%%%%%%%%%%%%%%
%%%%%%%%%%%%%%%%%%%%%%%%%%%%%%%%%%%%%%%%%%%%%%%%%%%%%%%%%%%%%%%%%%%%%%%%%%
%%%%%%%%%%%%%%%%%%%%%%%%%%%%%%%%%%%%%%%%%%%%%%%%%%%%%%%%%%%%%%%%%%%%%%%%%%

\clearpage
\onecolumngrid
\setcounter{equation}{0}
\setcounter{figure}{0}

\begin{center}
{\bf \Large 
Supplemental Material to\vspace*{0.3cm}\\ 
\emph{Dynamics of symmetry breaking during quantum real-time evolution in a minimal model system}
}
\end{center}

\vspace*{0.3cm}
{\center{
\hspace*{0.1\columnwidth}\begin{minipage}[c]{0.8\columnwidth}
In this supplemental material we provide methodological details for the flow equation renormalization group procedure used in the main text and for the calculation of observables. We determine the scaling equations for the couplings in the Hamiltonian and their analytical solutions in the relevant parameter regimes. Exploiting the exponential structure of the diagonalizing transformation of the Hamiltonian we introduce a novel scheme for the calculation of observables within the flow equation framework.
\end{minipage}
}
}

\section{A: Flow equations for the ferromagnetic Kondo model with magnetic field}
\label{app:uncertainty} 

The flow equation technique is a renormalization group procedure that aims at diagonalizing a weakly perturbed Hamiltonian successively through a sequence of infinitesimal unitary transformations~\cite{Kehrein2006}. For this purpose it provides a prescription in terms of an operator differential equation
\be
	\frac{d H(B)}{dB} = \left[ \eta(B),H(B) \right]
\ee
with $\eta(B)$ the anti-hermitian generator of the unitary transformation
\be
	U(B) = \mathcal{T}_B \exp \left[ \int_0^B dB' \,\, \eta(B') \right]
\label{eq:UBExponential}
\ee
where $\mathcal{T}_B$ denotes $B$-ordering analogous to common time-ordering with $\mathcal{T}_B [\eta(B)\eta(B')] = \eta(B')\eta(B)$ if $B'>B$, for example. At the end of the flow $B\to \infty$ the transformation $U(B)$ approximately diagonalizes  the Hamiltonian such that
\be
	H(B\to \infty) = U^\dag(B\to\infty) H U(B\to\infty)
\ee
is exactly solvable. Let $H=H_0+H_p$ denote the Hamiltonian with $H_0$ an exactly solvable part and $H_p$ the weak perturbation. Then, the choice
\be
	\eta(B) = \left[H_0(B),H_p(B) \right]
\ee
achieves this goal as long as one does not run into a strong-coupling divergence. For the ferromagnetic Kondo model with magnetic field $H = H_0 + H_p$ we have
\begin{align}
	H_0(B) & =  \sum_{k,\sigma = \uparrow,\downarrow}  \varepsilon_k c_{k\sigma}^\dag c_{k\sigma} - h(B)S^z , \nonumber \\ H_p(B) & = \sum_{kk'}  \left[ \frac{J^\ua_{kk'}(B)}{2} c_{k\ua}^\dag c_{k'\ua} - \frac{J^\da_{kk'}(B)}{2} c_{k\da}^\dag c_{k'\da} \right] S^z  +   \sum_{kk'}\frac{J^\perp_{kk'}(B)}{2} \left[ c_{k\ua}^\dag c_{k'\da} S^- + c_{k\da}^\dag c_{k'\ua} S^+ \right]
\end{align}
where for full generality we introduced a momentum dependence and anisotropy in the couplings which although not present in the initial model is generated during the flow. Following the usual prescription one obtains for the generator:
\begin{align}
	\eta(B) & = \eta^\parallel(B) + \eta^\perp(B), \quad
	\eta^\parallel(B) = \sum_{kk'} \Big[ \epsilon_k -\epsilon_{k'}  \Big] \left[ \frac{J^\ua_{kk'}(B)}{2} c_{k\ua}^\dag c_{k'\ua} -\frac{J^\da_{kk'}(B)}{2}c_{k\da}^\dag c_{k'\da} \right] S^z, \nonumber \\
	& \eta^\perp(B) = \sum_{kk'}\frac{J^\perp_{kk'}(B)}{2} \Big[ \epsilon_k -\epsilon_{k'} + h(B)  \Big] \Big[ c_{k\uparrow}^\dag c_{k'\downarrow} S^- -c_{k\downarrow}^\dag c_{k'\uparrow} S^+  \Big],
\label{eq:generator}
\end{align}
In the diagonal parametrization~\cite{Kehrein2006,Fritsch2010}
\be
	J^{\ua \da}_{kk'}(B) = j^{\ua \da}_{\overline{kk'}} e^{-B(\varepsilon_k-\varepsilon_{k'})^2},\quad J^\perp_{kk'}(B) = j^\perp_{\overline{kk'}} e^{-B(\varepsilon_k-\varepsilon_{k'}+h(B))^2}
\ee
with $\overline{kk'} = (\varepsilon_k+\varepsilon_{k'})/2$ this yields the following scaling equations for the couplings and the magnetic field in the thermodynamic limit:
\begin{align}
	\frac{dh(B)}{dB} = &\,\frac{1}{2} \int d\varepsilon d\varepsilon' [\varepsilon-\varepsilon'+h(B)] \left( g^\perp_{\frac{\varepsilon+\varepsilon'}{2}}\right)^2 \left[ n(\varepsilon) + n(\varepsilon') - 2n(\varepsilon) n(\varepsilon')  \right] e^{-2B(\varepsilon-\varepsilon'+h(B))^2} \nonumber \\
	\frac{dg^\ua_\varepsilon}{dB} = &- \int d\varepsilon' \, \tanh(\varepsilon'/(2T)) [\varepsilon-\varepsilon'+h(B)] \left( g^\perp_{\frac{\varepsilon+\varepsilon'}{2}}\right)^2 e^{-2B(\varepsilon-\varepsilon'+h(B))^2}\nonumber \\
	\frac{dg^\da_\varepsilon}{dB} =& - \int d\varepsilon' \, \tanh(\varepsilon'/(2T)) [\varepsilon-\varepsilon'-h(B)] \left( g^\perp_{\frac{\varepsilon+\varepsilon'}{2}}\right)^2 e^{-2B(\varepsilon-\varepsilon'-h(B))^2} \nonumber \\
	\frac{dg^\perp_\varepsilon}{dB} = &- \frac{1}{2} \int d\varepsilon' \, \tanh(\varepsilon'/(2T)) [\varepsilon-\varepsilon'-h(B)/2]  g^\ua_{\frac{\varepsilon + \varepsilon'-h(B)/2}{2}}g^\perp_{\frac{\varepsilon+\varepsilon'+h(B)/2}{2}} e^{-2B(\varepsilon-\varepsilon'-h(B)/2)^2}  \nonumber \\
	&- \frac{1}{2} \int d\varepsilon' \, \tanh(\varepsilon'/(2T)) [\varepsilon-\varepsilon' +h(B)/2]  g^\ua_{\frac{\varepsilon + \varepsilon'+h(B)/2}{2}}g^\perp_{\frac{\varepsilon+\varepsilon'-h(B)/2}{2}} e^{-2B(\varepsilon-\varepsilon'-h(B)/2)^2} 
\end{align}
see Ref.~\cite{Fritsch2010} where these equations have been derived in the context of the antiferromagnetic Kondo model. The only difference to the ferromagnetic case is the initial condition for the couplings $g$ which now are negative. Note that we have introduced the dimensionless couplings $g_\varepsilon = \rho j_\varepsilon$ as in the main text. The noninteracting density of states $\rho$ is chosen constant within the energy band $[-D,D]$ because we are only interested in the universal low-energy behavior. At low energies these scaling equations can be solved approximately analytically. At zero temperature for $B<h^{-2}$ one obtains by identifying $g(B) := g_0 (B)$~\cite{Fritsch2010}
\be
	h(B) = h \left[ 1 + \frac{g}{2}- \frac{g(B)}{2}  \right] ,\quad g^\perp(B) = g^{\ua\da}(B) = g(B) =   \frac{g}{1-g \log(\sqrt{B}D)}.
\ee
For $B>h^{-2}$ the flow stops within the current 1-loop accuracy~\cite{Fritsch2010} such that the final renormalized couplings read
\be
	h_\ast = h(h^{-2}),\quad g^\perp_\ast = g^{\ua\da}_\ast = g(h^{-2}) 
\ee
A nonzero temperature $T>0$, but still small enough such that $h\gg T$, will result in the identical flow equations within the present accuracy with only subleading corrections that can be neglected. This changes for temperatures $T>h$ where the flow essentially stops already at $B=T^{-2}$ yielding
\be
	h_\ast = h(T^{-2}), \quad g^\perp_\ast = g^{\ua\da}_\ast =g(T^{-2}).
\ee

\section{B: Local observables via cumulant expansions}

In the following, we introduce a scheme for the calculation of observables which does not rely on the solution of an additional set of scaling equations for the operator of interest as is usually done within the flow equation framework, but is rather based on utilizing the specific exponential structure of the diagonalizing unitary transformation $U(B)$ in Eq.~(\ref{eq:UBExponential}). Determining expectation values of operators, here we take $S^z$, requires an efficient representation of $U(B) S^z U(B)^\dag$. Due to the commutation relation $[\eta^\parallel(B),S^z]=0$ and $\eta^\perp(B) S^z=-S^z \eta^\perp(B)$ we have that
\be
	U(B) S^z =  S^z\tilde{U}(B) , \quad \tilde{U}(B) = \mathcal{T}_B e^{\int_0^B dB[\eta^\parallel (B)-\eta^\perp(B)]}.
\ee
as an exact result. Using this identity the dynamics of the magnetization $\langle S^z(t) \rangle$ can be obtained by evaluating:
\be
	\langle S^z(t) \rangle = \mathrm{Tr} \left[ \rho \, e^{iHt} S^z e^{-iHt} \right] = \mathrm{Tr} \left[\, \rho\, S^z\, \tilde{U}^\dag \tilde{U}(t) U^\dag(t) U \, \right]
\ee
with $U = U(B\to\infty), \tilde{U} = \tilde{U}(B\to\infty)$, and $U(t) = e^{iH_0t} U e^{-iH_0t}$. The initial state $\rho = \rho_S \otimes \rho_{B}$ of our specific protocol is always chosen to factorize the quantum dot system $\rho_S$ and the fermionic bath $\rho_B$. Splitting the trace $\mathrm{Tr}=\mathrm{tr}_S \mathrm{tr}_B$ into the traces over the spin degree of freedom ($S$) and the fermionic bath ($B$) the above expression can be evaluated via an operator cumulant expansion~\cite{Kubo1962xh} with respect to the bath degrees of freedom, i.e., tracing out the bath,
\be
	\langle S^z(t) \rangle = \mathrm{tr}_S \left[ \rho_S \, S^z \, e^{\hat{f}(t)}\right] ,\quad \hat{f}(t) = 2\mathrm{tr}_B \left[ F(t)^2 \right],\quad F(t)=\int_0^\infty dB \, \left[  \eta^\perp(B,t)-\eta^\perp(B) \right].
\ee
This is the result including all contributions up to second order in the coupling strength. Using Eq.~(\ref{eq:generator}) one obtains
\begin{equation}
	\hat{f}(t) = - \int d\varepsilon \int d\varepsilon'\, \mathcal{J}^2_{\varepsilon \varepsilon'}\, N_{ \varepsilon \varepsilon}(S^z)\, \Big[ 1-\cos(\varepsilon+h_\ast)t \Big],
\label{eqfInt}
\end{equation}
with
\be
	\mathcal{J}_{\varepsilon\varepsilon'} = \int_0^\infty dB \, g^\perp_\varepsilon(B) \left[ \varepsilon' + h(B) \right] e^{-B(\varepsilon'+h(B))^2}, \quad N_{\varepsilon\varepsilon'}(S^z) = n\left(\varepsilon - S^z \varepsilon' \right)\left[1-n\left(\varepsilon - S^z \varepsilon' \right) \right]
\ee
where $n(\varepsilon)$ is the Fermi-Dirac distribution. Note that $\hat{f}(t)$ in the present case only depends on the spin operators via the projection $S^z$ onto the $z$-axis. Thus, we can write
\be
	\langle S^z(t) \rangle = \frac{1}{2} \Big[ e^{f_\ua(t)} - e^{f_\da(t)}\Big],\quad f_{\sigma}(t) = \langle \sigma |\hat{f}(t) |\sigma \rangle, \quad \sigma = \ua,\da,
\ee
where $e^{f_\sigma(t)}$ can be associated with the local magnetization when the system is not initialized in a superposition of $\ua$ and $\da$ states but rather only in $\sigma=\ua,\da$.

\section{C: Analytical asymptotics of the magnetization}

In this part of the supplemental material we obtain the analytical asymptotics of the integral in Eq.~(\ref{eqfInt}). For large times $t\gg D^{-1}$ only the low-energy excitations in the vicinity of the Fermi surface are important such that one can approximate $g^\perp_\varepsilon(B) \approx g^\perp(B)$. Then it is possible to perform the energy integrations yielding at zero temperature for $t\ll t_h$
\be
	f_\sigma(t) = f_\ast = -\int_{D^{-2}}^{h^{-2}}dB \int_{D^{-2}}^{h^{-2}} dB' \,\, \frac{g^\perp(B) g^\perp(B')}{2(B+B')^2} \Theta\left(h^{-2}- B-B' \right)
\ee
where $\Theta(x)$ is the Heaviside step function. Evaluating the $B$-integration perturbatively in $g$ and $h$ one obtains
\be
	f_\ast = -g^2 \left[ \frac{3}{2} + \frac{1}{1+g\log(h/D)} \right]
\ee
For the long-time dynamics for times $t$ beyond $t_h$ it is suitable to perform the following substitution
\be
	x=(\varepsilon + h_\ast)t,\,\, y=(\varepsilon + h_\ast) t, \,\, b = B/t^2,\,\, b' = B'/t^2.
\ee
In the zero temperature case we have to distinguish the two different spin contributions $\sigma=\ua \da$. While $f_\ua(t)=f_\ast$ even for times $t>t_h$ the opposite spin contribution $f_\da(t)$ will acquire a different dynamical behavior. The dominant contribution comes from $b,b'>1$ yielding after the energy integrations
\be
	f_\da(t) = -\frac{3}{8} \sqrt{\pi} h g_\ast^2 t \int_{1}^\infty db \int_1^\infty db' \,\, \frac{1}{(b+b')^{5/2}}
\ee
such that
\be
	f_\da(t) \stackrel{t\gg t_h}{\longrightarrow} -\sqrt{\frac{\pi}{8}} g_\ast^2 h t
\ee
In the presence of temperatures $0<T<h$ smaller than the magnetic field the $f_\ua(t)$ shows some further dynamics. After performing the energy integrations one obtains
\be
	f_\ua(t) = -\frac{\sqrt{\pi}}{4} g_\ast^2 h \frac{e^{-h/T}}{1-e^{-h/T}} t \int_0^\infty db \int_0^\infty db' \,\, \frac{e^{-1/4(b+b')}-2(b+b')\left[e^{-1/4(b+b')}-1 \right]}{(b+b')^{5/2}}.
\ee
The remaining integrals can be performed analytically yielding
\be
	f_\ua(t) \stackrel{t\gg t_T}{\longrightarrow}-\pi g_\ast^2 h \frac{e^{-h/T}}{1-e^{-h/T}} t
\ee

\end{document}